# Ultralow Thermal Conductivity and Thermoelectric Properties of Bi$_4$GeTe$_7$ with an Intrinsic van der Waal Heterostructure

Niraj Kumar Singh, Ankit Kashyap and Ajay Soni*

*School of Basic Sciences, Indian Institute of Technology Mandi, Mandi, Himachal Pradesh, India-175075*

*Author to whom correspondence should be addressed: ajay@iitmandi.ac.in

Abstract:

Ternary chalcogenides, having large crystalline unit cell and van der Waal stacking of layers, are expected to be poor thermal conductors and good thermoelectric (*TE*) materials. We are reporting that layered Bi$_4$GeTe$_7$, with alternating quintuplet-septuplet layers of Bi$_2$Te$_3$ and Bi$_2$GeTe$_4$, has an ultralow thermal conductivity, $\kappa_{total}$ ~ 0.42 Wm$^{-1}$K$^{-1}$ because of high degree of anharmonicity as estimated from large Grüneisen parameter ($\gamma$ ~ 4.07) and low Debye temperature ($\theta_d$ ~ 135 K). The electron dominated charge transport has been realized from the Seebeck coefficient, $S$ ~ - 82 µV/K, at 380 K, and Hall carrier concentration of $n_e$ ~ 9.8 x 10$^{19}$ cm$^{-3}$ at 300 K. Observation of weak antilocalization (*WAL*), due to spin-orbit coupling (*SOC*) of heavy Bi and Te, advocate Bi$_4$GeTe$_7$ to be a topological quantum material also. The cross-sectional transmission electron microscopy images show the inherent stacking of hetero-layers, which are leading to a large anharmonicity for poor phonon propagation. Thus, being a poor thermal conductor with a *TE* figure of merit, *ZT* ~ 0.24, at 380 K, the Bi$_4$GeTe$_7$ is a good material for *TE* applications.

Keywords: Layered chalcogenides, Ultralow thermal conductivity, Thermoelectricity, Topological Transport, Bi$_4$GeTe$_7$.

Thermoelectric materials are important for harvesting waste thermal energy by directly converting heat into useful electricity.[1] The efficiency of *TE* materials are governed by the dimensionless figure of merit, $ZT = \frac{S^2\sigma}{\kappa_{total}}T,$ where *S* is Seebeck coefficient, *σ* is electrical conductivity, *T* is absolute temperature and $\kappa_{total}$ denotes the total thermal conductivity which has contribution from electrons ($\kappa_e$) as well as lattice ($\kappa_l$).[1] The involved physical parameters, except $\kappa_l$, are intricately interdependent and thus limiting the usage of *TE* materials for commercial applications. After understanding the challenges of optimizing the charge carriers in *TE* research, recently, materials with intrinsically ultralow $\kappa_l$ have grabbed the scientific attention since it allows to modulate the *σ* with negligible rise in $\kappa_e$, independently.[2-5] In general, materials having high degree of anharmonicity shows ultralow $\kappa_l$ due to damping effects of phonon-phonon interactions, which plays an important role in engineering of their thermal properties.[5]

In *TE* materials with large unit cells, complex and layered crystal structures, the presence of large number of defects give a high degree of crystal anharmonicity hence reducing $\kappa_{total}$.[6, 3, 5] By adopting various methodologies such as nanostructuring, second phase nanoinclusions, mass contrast, soft phonons and phonon-glass/liquid-electron-crystal (*PGEC or PLEC*), a further reduction of $\kappa_l$ can also be achieved.[3, 7-13] In addition, the heterostructured *TE* materials have shown a promise for significant enhancement in *ZT* through (i) reduction of $\kappa_{total}$ from enhanced phonon scattering and (ii) enhancement of *S* from selective filtering of low energy charge carriers, simultaneously.[14] In literature, a *ZT* ~ 2.4 has been reported in the thin film heterostructure of $Bi_2Te_3$/$Sb_2Te_3$ through controlled manipulation of electrons and phonons.[15] A significant enhancement of the *S* through low energy charge carrier filtering have been reported for $Bi_2Te_3$/Te heterostructured thin films.[16] However, all these heterostructured *TE* compounds are artificially grown or chemical modified, thus required expensive and tedious ways for large scale productions.

Owing to their layered crystal structure and high degree of anharmonicity, the ternary chalcogenides, such as $Bi_2Se_2Te$, $Bi_2Te_2(Se/S)$, $Bi_2GeTe_4$, $Bi_4PbTe_7$, $Bi_4SnTe_7$, and $Pb_2Bi_2Se_5$ possesses ultralow $κ_{total}$.[5, 17, 18] With the narrow band gaps and layered crystal structures, the family of $(Bi/Sb)_2(Te/Se/S)_3$ chalcogenides is known for the excellent *TE* properties and first generation topological insulators (*TI*) also.[1, 7] The exotic surface states are arising because of the involvement of SOC of heavier metal like Bi/Sb/Pb/Sn, and Te.[19-22] In this regard, $Bi_4GeTe_7$ is an important *TE* as well as TI material with inherent alternate repeating quintuplet-septuplet heterostructured crystal structure. While, the magneto transport studies show the presence of *WAL* (T ≤ 5K) and supports possibility of topological surface states, the existence of high degree of crystal anharmonicity and large Grüneisen parameter, $γ$ ~ 4.07, leading to an ultralow value of $κ_{total}$ ~ 0.42 $Wm^{-1}K^{-1}$ at 380 K. Our results provide subtle understanding of the structural aspects for $Bi_4GeTe_7$ as a potential *TE* material.

Polycrystalline $Bi_4GeTe_7$ was grown by a solid-state melting route using high purity chunks of Bi (99.99 %), Ge (99.999%) and Te (99.99%) in the stoichiometric ratio. The vacuum sealed (~ $10^{-5}$ mbar) quartz tube, with pure elements, was heated to ~ 1123 K for ~ 24 h followed by water quenching. The obtained ingot was finely ground and pelletized into a ~ 10 mm dia. pellet, which was further vacuum sealed and subjected to an annealing at ~ 773 K for ~ 100 h. The annealing was done to improve the consolidation and homogenization of the compound. The annealed pellet was cut into a rectangular bar (2 x 2 x 10 $mm^3$) for thermoelectric transport measurements. Structural and phase analysis of the powder was carried out with a rotating anode Rigaku Smartlab X-ray diffractometer using $CuK_α$ radiation. High resolution transmission electron microscopy (HRTEM) was done using FEI make Tecnai G2 20 S-TWIN transmission electron microscope. Morphology and elemental mapping were identified using field-emission scanning electron microscopy (FESEM, Nova Nano SEM 450). Temperature dependent thermoelectric and magnetotransport properties were performed in a physical properties measurement system (PPMS, DynaCool, Quantum Design).

X-ray diffraction pattern of the finely ground powder of the annealed $Bi_4GeTe_7$ (Fig. 1(a)) have been indexed to Rhombohedral geometry (*space group* $P\bar{3}m1$, 164).[23] Rietveld refinement confirms $Bi_4GeTe_7$ as the primary phase (~ 96.5 %) with the inherent $Bi_2GeTe_4$ is identified as the secondary phase (~ 3.5 %).[24] The observation of secondary phase is quite likely due to the shorter annealing time during the synthesis.[25, 26] The refined lattice parameters for $Bi_4GeTe_7$ are $a = b = 4.36$ Å and $c = 23.92$ Å and the corresponding unit cell volume, $V \sim 393.64$ Å$^3$.[27] A schematic in Fig. 1(b), depicts the natural arrangements of alternating quintuplets ($Te^I$-$Bi$-$Te^{II}$-$Bi$-$Te^I$) and septuplets ($Te^I$-$Bi$-$Te^{II}$-$Ge$-$Te^{II}$-$Bi$-$Te^I$), which are separated by van der Waal gaps. The cross sectional HRTEM image in Fig. 1(c) clearly showcase the quintuplet-septuplet heterostructure arrangements in $Bi_4GeTe_7$. The global layered nature and hierarchical layered stacking is shown in SEM image (Fig. 1(d)).[5, 13] Elemental mapping performed on polished surface of the pellet (Fig. 2), suggests absence of isolated secondary phase of $Bi_2GeTe_4$ in $Bi_4GeTe_7$.

The large unit cell $Bi_4GeTe_7$ is also a topological quantum material because of the presence of heavy elements like Bi and Te, which provides adequate spin-orbit-coupling for observation of signature metallic surface states in magnetotransport studies.[19] Hall analysis suggest electrons as the dominating charge carriers with a concentration, $n_e \sim 9.8 \times 10^{19}$ cm$^{-3}$ at 300 K. The magnetoresistivity, $\rho_{xx}(B)$, is shown in Fig. 3(a), where a sharp dip near the zero-field region providing a clue for *SOC* driven *WAL*, below 5K.[20, 28-31] To understand further, we fitted the the magnetoconductivity data with Hikami-Larkin-Nagaoka (HLN) equation,

$$\Delta\sigma = \sigma(B) - \sigma(0) = -\frac{\alpha e^2}{2\pi^2\hbar}\left[\ln\left(\frac{\hbar}{4e\ell_\phi^2 B}\right) - \psi\left(\frac{1}{2} + \frac{\hbar}{4e\ell_\phi^2 B}\right)\right] + cB^2,$$

where $e$ is electronic charge, $\alpha$ is a dimensionless fitting parameter called prefactor, $B$ is applied magnetic field, $l_\phi$ is phase coherence length and $\Psi$ represents digamma function.[32, 33] From the fitting (Fig. 3(b)), the estimated phase coherence length $l_\phi$ decreases from ~ 161 nm (2.5 K) to ~ 67 nm (10 K), which suggest the inelastic scattering of electrons from phonons. Generally, single conducting surface channel correspond to the $\alpha$ ~ - 0.5,[34] however a significant contribution of bulk conductivity reduces prefactor, $\alpha$ from ~ - 22 (at 2.5 K) to ~ - 0.11 (at 10 K). Thus, the observation of WAL provides an evidence of involvement of bulk as well as topological surface states in $Bi_4GeTe_7$.

To understand the potential for TE applications, temperature dependence of the involved physical parameters, for ZT, have been studied in detail. The positive temperature coefficient of resistivity in $\rho_{xx}$ (T), indicates a metallic behavior or a narrow band gap degenerate nature of semiconductors, (Fig. 3(c)), with a residual resistivity ratio (*RRR*) of ~ 1.87.[13] A relatively higher $\rho_{xx}$ for $Bi_4GeTe_7$, in comparison with ternary chalcogenides shows a higher crystal complexity involved and an additional resistance to the electron flow in the material.[5, 35] The *S(T)*, in Fig. 3 (d), further confirms the electron dominated charge transport with $S$ ~ - 82 μV/K and correspondingly, the power factor (inset of Fig. 3(d)), $PF$ ~ 2.65 x $10^{-4}$ $Wm^{-1}K^{-2}$ at 380 K, has been determined.

The $\kappa_{total}$ shows ultralow values, throughout the temperature range, with a $\kappa_{total}$ ~ 0.42 $Wm^{-1}K^{-1}$ at 380 K (Fig. 4(a)), which assures that $Bi_4GeTe_7$ has a potential for better *TE* material. The contribution of $\kappa_e$ and $\kappa_l$ in $\kappa_{total}$, is estimated using the modified Wiedemann Franz law: $\kappa_e = \left[1.5 + exp\left[-\frac{|\alpha|}{116}\right]\right]\sigma T$.[5, 36] Thus $\kappa_e$ ~ 0.29 $Wm^{-1}K^{-1}$ and $\kappa_l$ ~ 0.13 $Wm^{-1}K^{-1}$, at 380 K, are extracted from temperature dependence of the Lorenz number, *L* as shown in the inset of Fig. 4(a). Here the interesting crystal structure of $Bi_4GeTe_7$, with quintuplet-septuplet heterostructure having van der Waal's gaps, has sufficient hindrance for phonon propagation as reflected from the ultralow values of $\kappa_l$. The room temperature comparison of ultralow values of $\kappa_l$ of various layered chalcogenides such as $Bi_2Te_3$, $Bi_2GeTe_4$, $GeBi_3SbTe_7$, $GeBi_2Sb_2Te_7$, $GeBiSb_3Te_7$ and $Sb_4GeTe_7$ is shown in Fig. 4(b), which showcases the importance of structural complexies of quintuplet-septuplet heterostructure in $Bi_4GeTe_7$.[5, 13, 35]

In *TE* materials, the higher crystalline anharmonicity drives phonon relaxation processes by damping the exchange of energy between thermal and mechanical vibrations, and thus resulting in ultralow $\kappa_l$.[5, 6, 37] In order to understand the origin of ultralow $\kappa_l$, the anharmonicity can be assessed from the heat capacity data by estimating the Debye temperature ($\theta_d$) and the Gruneisen parameter ($\gamma$), of the materials. The low temperature heat capacity ($C_p$) data is shown in Fig. 4(c)), and the linear fit (inset of Fig. 4 (c)), of $C_p/T = A + BT^2$,[38] gives the characteristic constants $A$ (~ 1.81 x $10^{-3}$ J $mol^{-1}$ $K^{-2}$) related to electronic contributions and $B$ (~ 9.31 x $10^{-3}$ J $mol^{-1}$ $K^{-4}$) related to lattice contributions in $Bi_4GeTe_7$. The $\theta_d$ ~ 135.7 K has been estimated using, $\theta_d = \left(\frac{12\pi^4 pR}{5B}\right)^{1/3}$, where $p$ is the number of atoms per formula unit and $R$ is the molar gas constant.[17] The Grüneisen parameter, $\gamma$, is calculated using the relation: $\kappa_l = \frac{P\bar{M}\delta\theta_d^3}{\gamma^2 N^{2/3} T}$, where $\bar{M}$ is the averaged mass of atoms of the material, $\delta^3$ is the volume per atom, $N$ represents the total number of atoms in the primitive unit cell and $P = 3.1$ x $10^{-6}$ is a constant (the units of $\delta$ is in Å, $\kappa_l$ is in $Wm^{-1}K^{-1}$ and $\bar{M}$ is in amu).[5, 17, 39] Here, $Bi_4GeTe_7$ possesses a high value of $\gamma$ ~ 4.07 (at 380 K), which is better than champion $Bi_2Te_3$ and comparable to the *TE* materials known for ultralow thermal conductivity, such as $Bi_2GeTe_4$, $Ag_8SnSe_6$ and $Ag_8GeSe_6$.[3, 5] The low $\theta_d$ and large $\gamma$ signify that $Bi_4GeTe_7$ has a high degree of anharmonicity. An additional insight on the nature of phonon propagation in $Bi_4GeTe_7$ can be drawn from the calculation of average phonon velocity, ($v_{avg}$). In general, materials with stiffer bonds tends to have higher $v_{avg}$ and hence higher $\kappa_{total}$, whereas smaller $v_{avg}$ corresponds to lower $\kappa_{total}$. For Bi$_4$GeTe$_7$, $v_{avg}$ ~ 1.49 km s$^{-1}$, has been estimated using the equation $v_{avg} = \frac{2\pi k_B \theta_d}{(6\pi^2 n_v)^{1/3} h}$, where $k_B$ and h represents Boltzmann constant and Planck's constant, respectively, while $n_v$ denotes number of atoms per unit volume of the material. The $v_{avg}$ for Bi$_4$GeTe$_7$ is considerably smaller when compared to ultralow $\kappa_{total}$ materials such as Bi$_2$GeTe$_4$ ($v_{avg}$ ~ 1.54 km s$^{-1}$) and Bi$_2$Te$_3$ ($v_{avg}$ ~ 1.78 km s$^{-1}$).[5, 17] Therefore, intrinsically ultralow $\kappa_{total}$ in Bi$_4$GeTe$_7$ can be understood as a result of anharmonicity arising due to large unit cell, intrinsic heterostructures, presence of heavy Bi atoms, large $\gamma$ and small $v_{avg}$. Furthermore, the calculated *ZT* at 380 K is ~ 0.24 (Fig. 4(d)), which is relatively higher than the similar ternary layered chalcogenides.[5, 35] The results establish that Bi$_4$GeTe$_7$ has potential for good *TE* Material. Moreover, other members of Bi based ternary layered chalcogenides with higher structural complexities such as (Bi/Pb)$_6$GeTe$_{10}$, (Bi/Pb)$_2$GeTe$_5$ and (Bi/Pb)$_2$Ge$_3$Te$_6$ can be expected to show ultralow $\kappa_{total}$ and host topological surface states which upon proper tuning of the electrical properties can be good *TEs* and *TIs*.

In summary, the structural aspects of Bi$_4$GeTe$_7$ has been discussed for its magneto transport, TE transport properties. Here, the possibility of topological surface states have been understood by the $l_\phi$ ~ 161 nm (2.5 K) and an existence of WAL in magneto transport studies. With *S* ~ - 82 µV/K at 380 K, and $n_e$ ~ 9.8 x 10$^{19}$ cm$^{-3}$ at 300 K, the Bi$_4$GeTe$_7$ has an electron dominated charge transport like degenerate semiconductors. The intrinsic van der Waal heterostructure of quintuplet-septuplet layers provide a large anharmonicity and phonon scattering for an ultralow $\kappa_l$ ~ 0.13 Wm$^{-1}$K$^{-1}$ at 380 K. Hence, Bi$_4$GeTe$_7$ is an important topological quantum material and promising *TE* material with high $\gamma$ ~ 4.07 and an overall *ZT* ~ 0.24 at 380 K.

NKS would like to acknowledge CSIR-SRF and AK acknowledges SERB for providing the fellowship. AS would like to acknowledge DST-SERB (Grant No. CRG/2018/002197). Authors thank IIT Mandi for the research facilities.

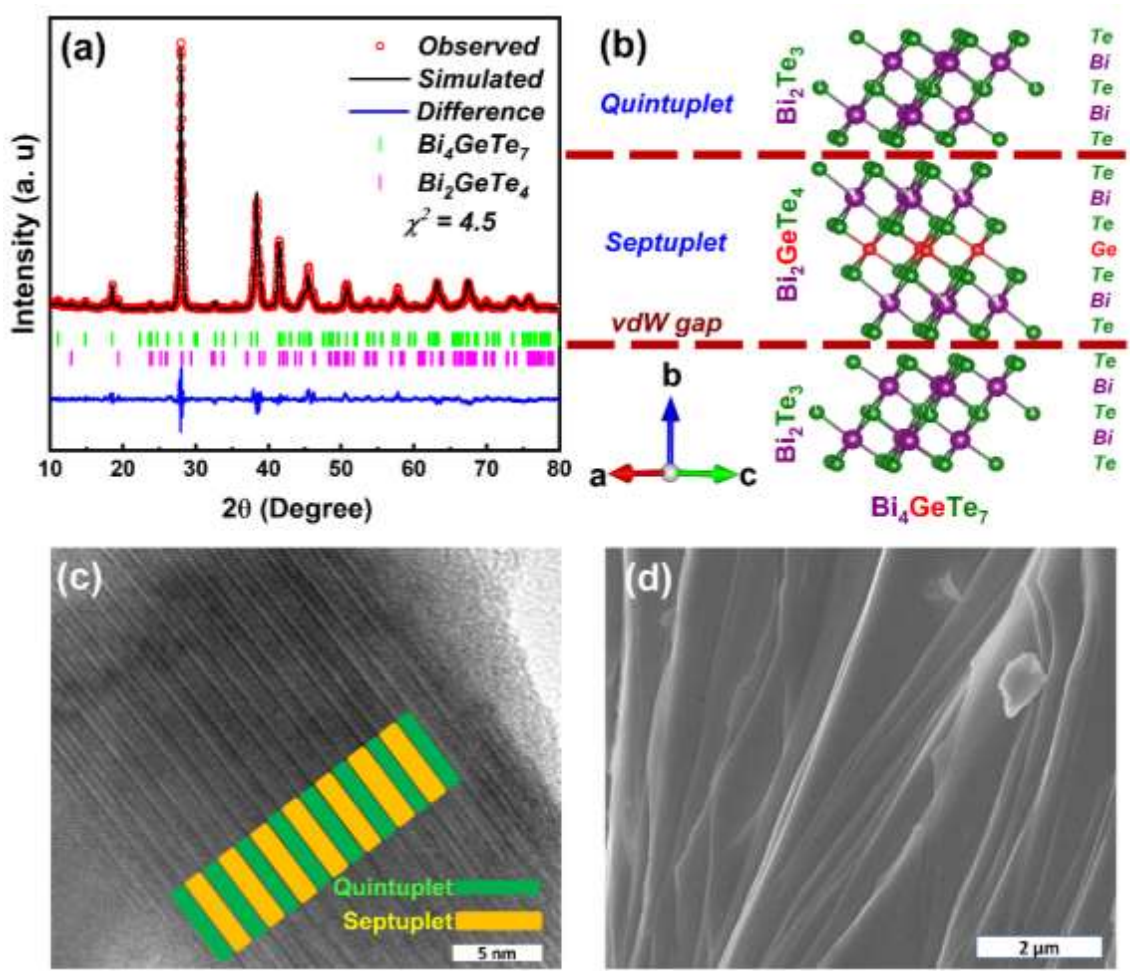

FIG 1. (a) Rietveld refinement of powder XRD pattern, (b) Schematic of the quintuplet-septuplet hetero-structure with vdW gaps of $Bi_4GeTe_7$, (c) cross-sectional TEM image showcasing the stacked layers at atomic scales, and (d) FESEM image of fractured crystal showcasing global pictures of the layers.

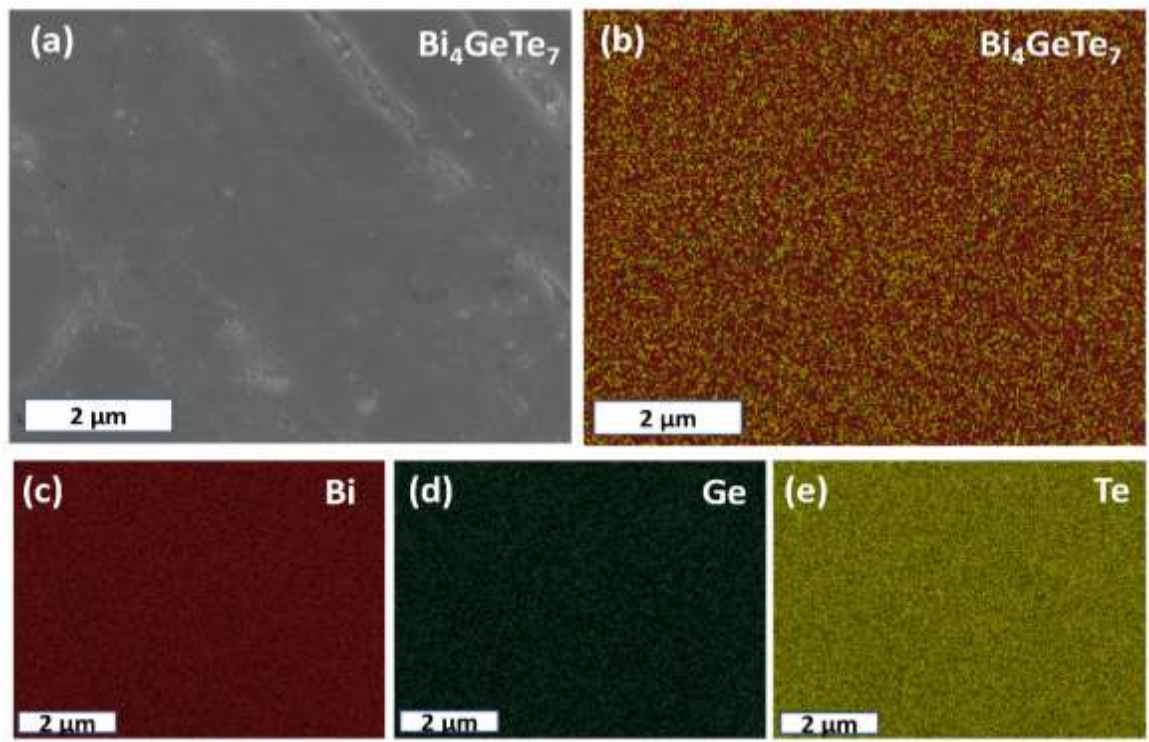

FIG 2. (a) FESEM image of top view of the pellet, and (b-f) the homogenous distribution of various elements observed in the mapping performed on polished surface of pellet.

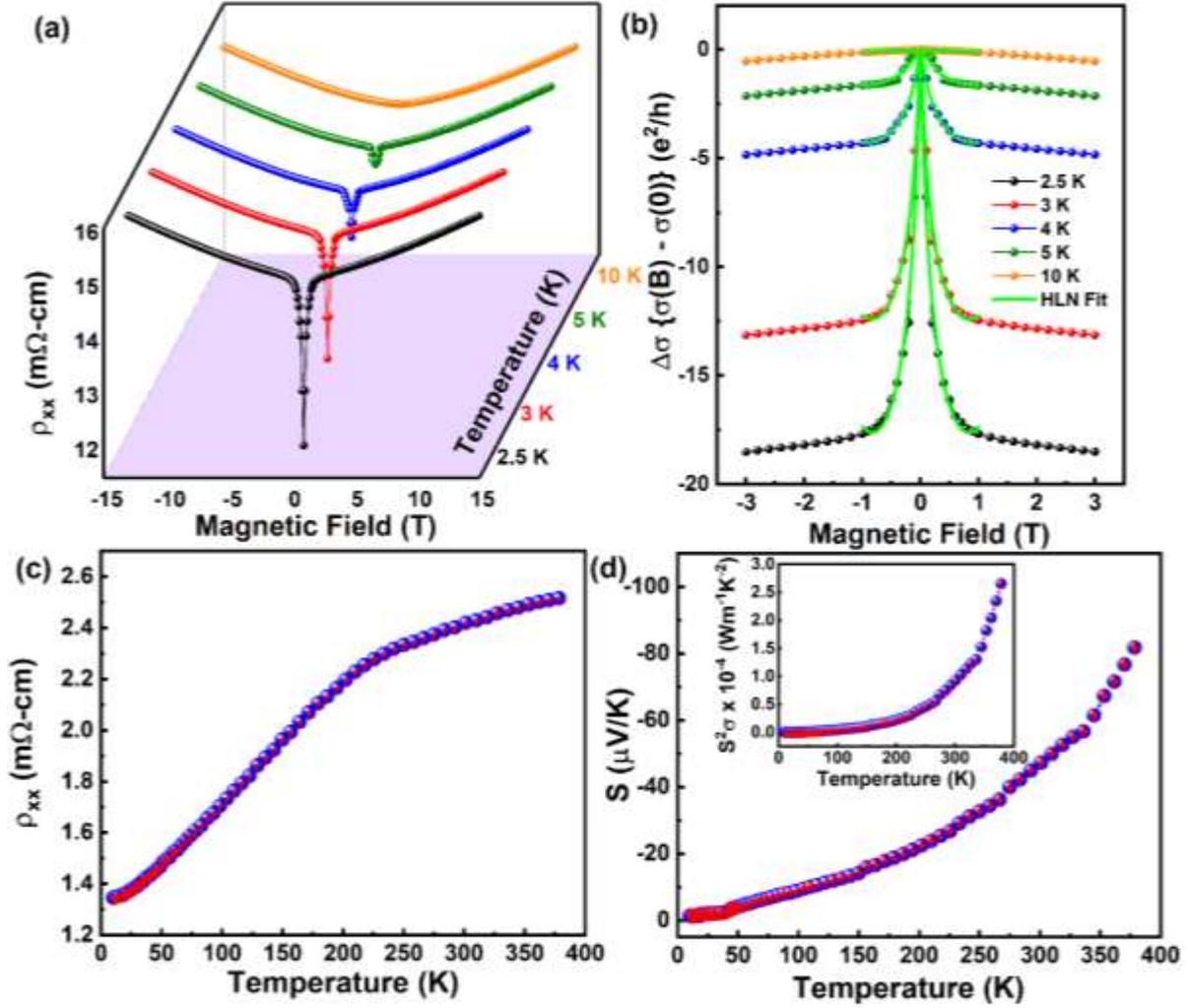

FIG 3. Magnetic field dependence of (a) longitudinal magnetoresistivity ($\rho_{xx}$), and (b) HLN fitted magnetoconductivity, and temperature dependence of (c) $\rho_{xx}(T)$, (d) $S(T)$ and (Inset $S^2\sigma$).

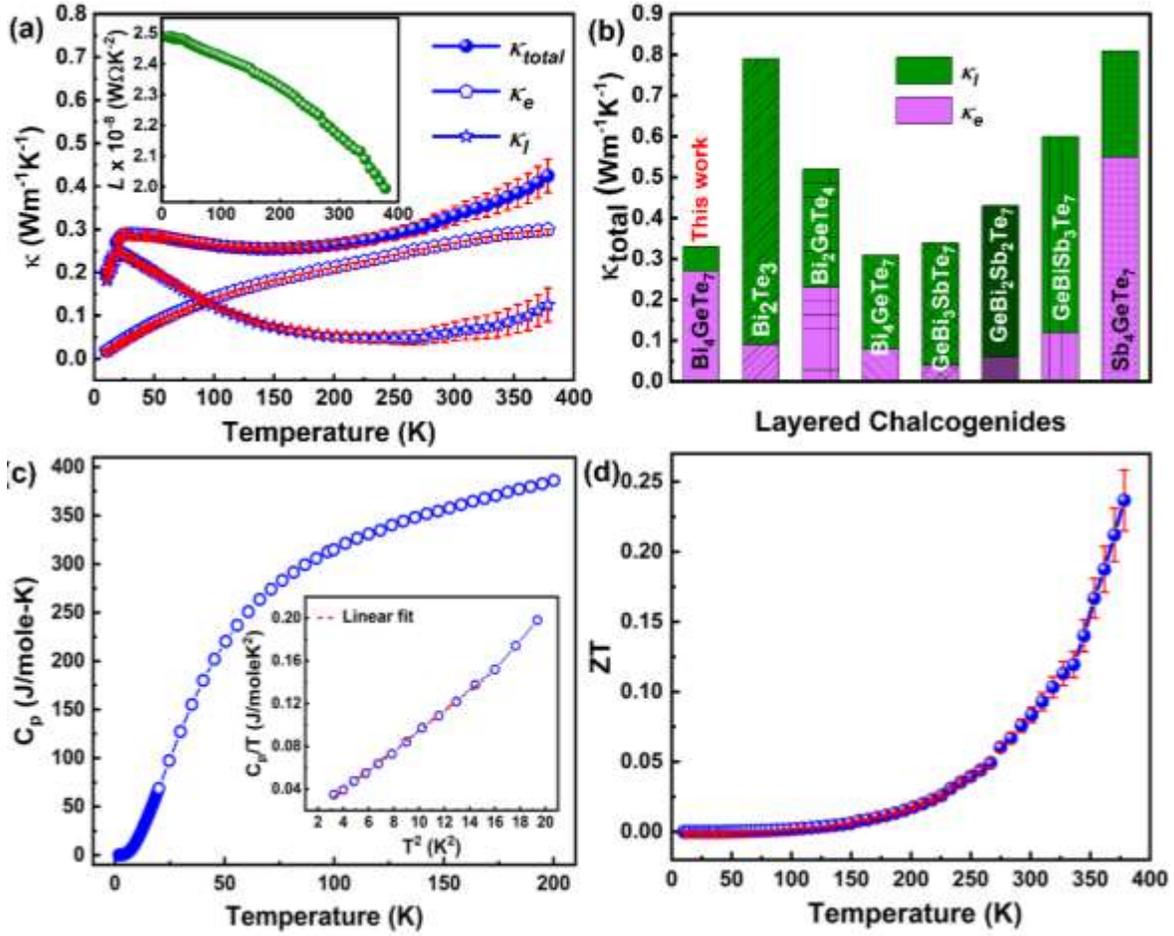

FIG 4. (a) Temperature dependence of $\kappa_{total}$, $\kappa_{el}$, $\kappa_l$, and inset shows $L(T)$), and (b) comparative graph for $\kappa_{total}$, $\kappa_{el}$ and $\kappa_l$ of various layered chalcogenides.[5, 13, 35] Temperature dependence of (c) $C_p$ (Inset: $C_p/T$ vs $T^2$) and (d) $ZT$.